# A stochastic approach to the filling dynamics of a porous medium: full/empty pores duality symmetry and the emergence of Darcy's law

Robert Bouzerar[1], Issyan Tekaya[1]

[1]Laboratoire de Physique de la Matière Condensée, Université de Picardie Jules Verne, 33 rue Saint Leu, 80039 Amiens Cedex, France

**Abstract.** A stochastic approach to the filling dynamics of an open topology porous structure permeated with a perfectly wetting fluid is presented. From the discrete structure of the disordered voids network with only nearest neighbors links, we derive the "microscopic" (at the pores scale) dynamical equations governing the filling dynamics of the coupled pores and the fluid pressure dynamics. The model yields two fundamental consequences. The first consequence regards the emergence of Darcy's law and the dependence of the predicted permeability with the voids network topology. The second one is the prediction of a diffusive dynamics for the degrees of freedom of the pores filling. These equations exhibit a new type of symmetry manifested by their invariance under the full/empty pores duality transformation jointly with the velocity reversal. Non-trivial steady non-equilibrium pores filling states are also obtained and found to follow a Fermi-Dirac type law. The analogy with the single occupation of lattice sites by fermions is highlighted together with the corresponding hole-particle symmetry.

**Keywords.** Porosity. Darcy's law. Stochastic process. Duality symmetry.

## 1. Introduction

Porous media mainly consist of a solid matrix hosting a more or less dense network of partially interconnected voids which can exhibit a great variety of geometrical structures, resulting in a tight coupling between the solid matrix and the fluid flowing through the structure. The investigation of such media is a cross-disciplinary field, falling within both solid mechanics and hydrodynamics. Their obvious complexity arises from the non-trivial topology (disordered voids network/connectivity features) which governs their puzzling physical behaviors. This topological complexity is also the main obstacle to overcome on the way to a comprehensive theory of these media.

The understanding of fluid transport in porous media is an old but challenging problem [1], partly because of their wide range of applications. Various laboratory or industrial applications involving porous structures such as membrane filtration, water flow through granular media, hydrocarbons exploitation or pollutants trapping will profit from its solution. But the more significant gain of such a physical understanding certainly regards the fundamental side where so many issues remain unanswered. Yet many difficulties obstruct the way to a satisfactory physical and mechanical solution to that problem. Among these difficulties, we must face the triple complexity of porous media: the influence of the topology of the interconnected voids network, including disorder, the coupling between the matrix elasticity and the fluid flow and lastly, the prominent role of the spatial correlations of the filling of the pores. This last issue is usually not addressed in the available studies in which the medium is saturated with the fluid.

Though not completely understood, the physical consequences of such complex topologies and geometries are highly varied. In porous media with an open topology (connected voids), the enhanced coupling between the solid matrix and the fluid phase due to their complex geometries generates significant variations of the velocity field (as well as thermal variations according to the velocities amplitudes) resulting in intense viscous dissipation. The propagation of acoustic waves (pressure variations) are also strongly affected by that singular geometry of porous media. A comprehensive account for that complexity is inaccessible and therefore, simplifications are necessary. Subsequently, the usual approaches to these many and various effects rely on a limited set of parameters capturing the relevant features of the complex geometry of porous media. Some of these parameters have a direct geometrical signification: the porosity $\phi = V_f/V$ is the volume fraction associated with the voids space (or the fluid if the medium is saturated with the fluid), the tortuosity [2,3] for arbitrary shape voids is a dimensionless parameter comparing the mean microscopic kinetic energy of any inviscid incompressible fluid flowing through the structure to its macroscopic kinetic energy. That parameter incorporates the variations of the pores diameters and accounts for the curvature variations of the pores (or of the fluid streamlines). These two parameters can be assessed from appropriate measurements. Additional parameters regarding the fluid can be defined, such as the specific resistance opposed to the fluid as deduced from Darcy's law [1,4], Darcy's permeability (effective section endowed with area units) which does not depend on the type of fluid and reflects the way the internal geometry of the porous medium affects the flow. This permeability can be assessed from frequency-dependent viscous losses within the medium. Other parameters are connected with thermal properties of such complex media.

Many studies [5–7], especially numerical simulations, pointing out the importance of the topological features, are based on structural models of the matrix. In such models the structural complexity comes down to two global quantities defined previously, the porosity/tortuosity associated with geometrical features and the permeability capturing the influence of the medium on fluid transport. Such models suit especially mechanical needs but meet difficulties to handle the coupling to the flow properties. Though the reasons for these failures are difficult to identify clearly, it seems that the tentative reduction to a continuum description (necessary to apply the usual laws of mechanics and fluid flow) through an effective continuum medium, could be responsible for such a situation. The notion of an effective medium is introduced through Terzaghi's principle [8]. This approach relies on Biot's consolidation theory [9,10] treating the hydromechanical coupling within porous structures by means of an effective stress tensor supported by the solid matrix and incorporating the influence of the fluid pressure $p_f$,

$$\sigma_{ij} = \sigma_{ij}^{eff} - \alpha M p_f \delta_{ij} \tag{1}$$

Similarly, the fluid pressure depends on the matrix deformation. The coupling is provided by Biot's modulus $M$ which captures the main features of the pores network (not described explicitly in this continuum approach) and leads to the renormalization of the elastic modulii of the solid matrix [11]. The mechanical equations derived from equation (1) are supplemented by Darcy's law describing the fluid transport,

$$\vec{v} = -\frac{k}{\mu}\vec{\nabla}p_f \qquad (2)$$

This is a macroscopic velocity field depending on both the fluid dynamic viscosity $\mu$ and the (nonlinear) permeability $k = k_0 e^{M\varepsilon}$ sensitive to the matrix deformation.

The study reported in this paper depicts an approach of porous media different from an effective continuum description. We propose a local approach to the fluid transport through random porous media based on a stochastic description of the flow exchange between neighboring pores. This probabilistic description incorporates the disordered voids network to account for the fluid diffusion through the structure.

## 2. Description of the structural model

The typical geometry of porous media we are interested in is schematized on figure 1. The pores (circles of the figure), with possibly varying sizes, are randomly distributed within the solid matrix. The fluid can flow between neighboring pores connected by small diameter channels. Given the fluid viscosity, the diameters and lengths of the channels determine their conductances ($\gamma_{ij}$). The relevant notion of state of the network is given here by the pressure values ($p_i$) and the matrix grouping the conductances. Many flowing regimes can be tested by an appropriate choice of conductances and pressure: static values correspond to the Poiseuille flow while modulated pressures can, for instance, account for the Womersley regime associated with frequency-dependent conductances.

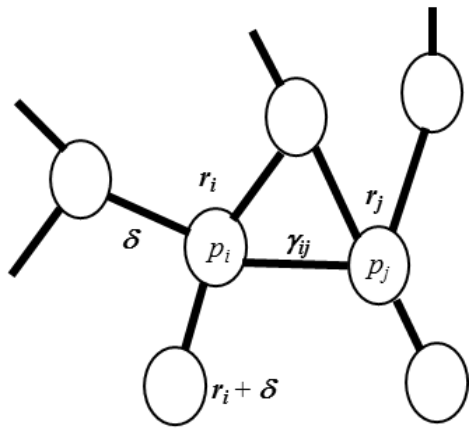

*Figure 1. Typical geometry of a disordered porous network with an open topology. That structure is encoded by a graph with the pores as summits connected by links associated with conductances $\gamma_{ij}$.*

The fluid pressure within the pores and the conductance matrix of the connections network allow to define a specific graph encoding the topology of the porous structure under consideration. We will show in the next section that this graph determines unambiguously the dynamical behavior of the fluid flowing through the structure. The mechanical properties of the solid matrix (rigid or slightly deformable) and their coupling to the flow dynamics can be incorporated to the associated graph structure through additional features of the pores as their compliances $C_i$. The static compliance of any (arbitrary shape) pore is defined as its volume variation (pore deformation) due to the pressure variation of the fluid within. In the presence of a modulated flow, it can be generalized as a dynamical compliance connecting the Fourier spectra of the volume variations and the pressure variations (within the approximation of linear response). It is clear that this notion accounts for the mechanical coupling between the pores and the solid matrix.

## 3. Coupling between pressure and filling dynamics

We are interested here in situations of only partial filling of the voids. The usual situation of a porous domain saturated with an incompressible fluid has been addressed in many complete studies. The degrees of freedom $\varepsilon$ of the pores filling are defined as the volume fraction of the pores that are filled with the fluid. We will restrict ourselves to the rigid matrix limit: in the situation of a complete filling of the pores, such a limit seems trivial (especially for an incompressible flow) and the overall dynamics is dominated by the fluid flow. But for partial filling, the situation is not trivial and requires the knowledge of the filling dynamics. Applying the mass conservation law to the graph of figure 1 yields for the *i-th* pore,

$$\frac{dv_i(t)}{dt} = -\sum_j \gamma_{ij}(p_i(t) - p_j(t)) \qquad (3)$$

The volume of the pore is $V_i(t) = \varepsilon_i(t)V$, $V$ being the invariable average volume of the empty pores (maximal fluid volume). As the sum runs over the nearest neighbors of pore $i$ with relative positions $\vec{r}_j = \vec{r}_i + \vec{\delta}$ indexed by the random vector $\vec{\delta}$. Its average length gives the mean separation between two neighboring pores (or equivalently it determines the average pores density within the matrix). The statistical distribution of the directions $\vec{\delta}$ around each pore (connectivity fluctuations) is characterized by its average $<\vec{\delta}> = \vec{0}$ and the correlation matrix $\Gamma$ with coefficients $\Gamma_{ab} = <\delta_a\delta_b>$. As a result, equation (3) becomes, after a power-series expansion (component of $\vec{\delta}$) truncated to the 2$^{nd}$ order,

$$V\frac{d\varepsilon_i(t)}{dt} \approx \frac{1}{2}z\gamma\vec{\nabla}.\left(\Gamma.\vec{\nabla}p\right)(\vec{r}_i, t) \qquad (4)$$

In this equation, we have introduced the average number $z$ of neighbors of any pore and have supposed a weak dispersion of the channels lengths so that the conductances, which depend only on their length, are almost constant. Introducing the pores density $n(\vec{r}) = \overline{\sum_i \delta(\vec{r} - \vec{r}_i)}$ averaged (overbar) over the positions of the pores, we are led to the reinterpretation of equation (4) derived from the global equation,

$$\frac{d}{dt}\left(\rho_f V \int_D \overline{\sum_i \varepsilon_i(t)\,\delta(\vec{r} - \vec{r}_i)}\,d^3\vec{r}\right) = \frac{1}{2}\rho_f z\gamma \int_D n(\vec{r})\,\vec{\nabla}.\left(\Gamma.\vec{\nabla}p(\vec{r}, t)\right)d^3\vec{r} \qquad (5)$$

The domain $(D)$ is the space domain containing the porous structure. The left-hand side member is the total fluid mass variation, the integrated expression being its mass density modified by the porous structure (the density of the "free" fluid being $\rho_f$). It follows from (5), treated as a mass conservation equation, that the effective velocity field of the fluid becomes,

$$\vec{v}_D(\vec{r},t) = -\frac{1}{2}z\gamma n(\vec{r})\Gamma.\vec{\nabla}p \tag{6}$$

We will naturally refer to the prediction (6) as the Darcy velocity field since it has the form of the usual Darcy's law. More precisely, equation (7) generalizes that law to anisotropic situations or equivalently to a permeability tensor,

$$k_{ab} = \frac{1}{2}z\gamma\mu n(\vec{r})\Gamma_{ab} \tag{7}$$

We have introduced the fluid proper viscosity $\mu$ related to the conductance by a relationship of the type $\gamma \propto \frac{1}{\mu}$. In the simpler situation of a homogeneous system (constant pore density) and isotropic that is, $\Gamma_{ab} = <\delta_a\delta_b> = <\vec{\delta}^2>\frac{\delta_{ab}}{3}$, we get the more suggestive expression of the permeability,

$$k_{ab} = \frac{z\gamma\mu n}{6}<\vec{\delta}^2>\delta_{ab} \tag{8}$$

In a disordered medium, the isotropic assumption is certainly natural in the absence of external stresses applied to the solid matrix, but a mechanical deformation of the medium can affect the correlation matrix. Such a more general situation will not be addressed in the present paper. The most obvious conclusion to be drawn from our study is the following one: our statistical approach leads rather naturally to Darcy's law generalized to inhomogeneous and anisotropic situations. It also predicts the dependence of the permeability upon the internal features of the medium (network connectivity, pores density, anisotropy due to the shape of the pores, etc.).

## 4. Filling dynamics: probabilistic approach, analogy with fermions

The last section emphasized the importance of the filling dynamics (Eq. 3,4). Nevertheless, the treatment of fluid transport through a porous structure as presented in the previous section, is not the general situation. It was assumed that the conductances were almost constant with a symmetrical conductance matrix. These simplifications are not realistic for a porous structure filled with an incompressible fluid. The fluid transfer between neighboring pores will depend on their filling state: fluid transfer to a saturated pore will not occur. To understand the dependence of the transport coefficient upon the filling variables, a more detailed analysis of the processes involved in fluid transport is required.

As both deterministic and stochastic processes (because of disorder) are present, a probabilistic approach is relevant. Let $p_i(1,t)$ be the probability for the $i$-th pore to be filled (busy) at any time $t$ and $p_i(0,t) = 1 - p_i(1,t)$ the corresponding probability for emptiness. These probabilities are associated with a set of two-valued random variable $e_i = 0$ or $1$. The average values of these variables are then,

$$<e_i>(t) = 1 \times p_i(1,t) + 0 \times p_i(0,t) = p_i(1,t)$$

It follows from that result that the filling variables $\varepsilon_i(t) = <e_i>(t) = p_i(1,t)$ can be identified as these probabilities. Their evolution proceeds from the master-equation governing the evolution of these probabilities, which reads,

$$\frac{dp_i(1,t)}{dt} = \sum_j [\omega_{ji} p_j(1,t)(1-p_i(1,t)) - \omega_{ij} p_i(1,t)(1-p_j(1,t))] \quad (9)$$

The coefficients $\omega_{ji}$ give the probability per unit time for the fluid to be transferred from pore $i$ to $j$. As stated previously, these symbols are not symmetric, because they depend on the pressure difference between pores $i$ and $j$. These symbols comprise two contributions,

$$\omega_{ij} \approx \omega_{ij}^0 + K_{ij}(p_i - p_j) \quad (10)$$

The first one $\omega_{ij}^0 = \omega_{ji}^0$ does not depend on the pressure difference and accounts for a spontaneous fluid transfer in the absence of an external pressure gradient (diffusion). The second one, non-symmetrical, is determined by the coefficients $K_{ij}$ (resp. $K_{ji}$) which should be non-zero only when $p_i > p_j$ (resp. $p_j > p_i$) so that it can be written $K_{ij} = K\theta(p_i - p_j)$ (resp. $K_{ji} = K - K_{ij}$) where $\theta$ designates the Heaviside function. This term introduces a very complex non-linearity in the model. Inserting this decomposition into Eq. (9) yields the filling factor dynamics,

$$\frac{d\varepsilon_i(t)}{dt} = \sum_j \omega_{ij}^0 (\varepsilon_j - \varepsilon_i) + \sum_j [K_{ji}\varepsilon_j(1-\varepsilon_i) - K_{ij}\varepsilon_i(1-\varepsilon_j)](p_j - p_i) \quad (11)$$

Comparing this equation to Eq. (3) leads to the identification:

$$\gamma_{ij} = V\left(K_{ji}\varepsilon_j(1-\varepsilon_i) - K_{ij}\varepsilon_i(1-\varepsilon_j)\right) \quad (12)$$

We are thus led to effective conductances depending on the filling states of the pores. For an empty network ($\varepsilon_i = 0$) these conductances are naturally vanishing. For pores saturated with the fluid, we also obtain zero, which agrees with the expected behavior of an incompressible fluid. For a uniform filling (< 1), we recover equation (3), with conductance dominated by its antisymmetric part. It can be easily shown from equation (11), that the first contribution to the filing dynamics is a diffusive process of the filling factor with a diffusion coefficient $D_0 = \omega^0 <\vec{\delta}^2> z/6$. Though more difficult to obtain, the most interesting consequences regards the second contribution. The equivalent continuum description of the discrete formulation defined by equation (11) can be obtained by introducing the average pore density $n(\vec{r}) = \overline{\sum_i \delta(\vec{r} - \vec{r}_i)}$ (see section 2) and integrating over an arbitrary domain $D$. It should be noticed that this continuum description is not the continuum limit of our discrete model, which would be valid only in the limit of high pore densities. Our equivalent continuum description amounts to the introduction of regular functions $\varepsilon$ (filling) and $p$ (pressure) which coincides with our variables when evaluated on the pores. We will present here only the result of that delicate procedure. We obtain,

$$\frac{\partial \varepsilon}{\partial t} = D_0 \Delta \varepsilon + \frac{1}{6} zK <\delta^2> [<\theta\left(p(\vec{r}+\vec{\delta})-p(\vec{r})\right)\varepsilon(\vec{r}+\vec{\delta})(1-\varepsilon(\vec{r}))> - $$
$$<\theta\left(p(\vec{r})-p(\vec{r}+\vec{\delta})\right)\varepsilon(\vec{r})\left(1-\varepsilon(\vec{r}+\vec{\delta})\right)>]\Delta p \qquad (13)$$

The brackets mean an average of the quantities over the directions $\vec{\delta}$. This equation can be regarded as the manifestation of the complexity of porous systems. It also illustrates the importance of the coupling between the filling and the pressure dynamics. Apart from that coupling, the main signature of the complexity relies on the filling space-correlations $<\varepsilon(\vec{r})\,\varepsilon(\vec{r}+\vec{\delta})>$, restricted to neighboring pores: these nearest neighbors filling correlations are a manifestation of short-range order in the network. For higher pores separation (low density), the filling states are statistically uncorrelated. This issue is rarely addressed in usual models of porous media. These correlation functions can be expanded as,

$$<\varepsilon(\vec{r})\,\varepsilon(\vec{r}+\vec{\delta})> \approx \varepsilon^2(\vec{r}) + \frac{1}{2} <\vec{\delta}^2> \varepsilon \Delta \varepsilon \qquad (14)$$

This expansion generates non-trivial filling/pressure coupling (gradient coupling, Laplacian coupling, etc.) in Eq. 13. This interesting consequence of the model will be analyzed further in next studies because of its potential importance in some applications (industrial or biological). Dropping these complex corrections, the simplest model we are led to is described by the equation,

$$\frac{\partial \varepsilon}{\partial t} \approx D_0 \Delta \varepsilon + \frac{1}{6} zK <\vec{\delta}^2> \vec{\nabla}.(\varepsilon(1-\varepsilon)\vec{\nabla}p \qquad (15)$$

Inserting in Eq. 15 the effective fluid density, $\rho_{eff} = nV\rho_f \varepsilon = N_p \phi \rho_f \varepsilon$, proportional to the porosity and the number of pores we get an effective current,

$$\frac{\vec{j}_{eff}}{\rho_f} = -D_0 N_p \phi \vec{\nabla}\varepsilon - \frac{1}{6} zKV <\vec{\delta}^2> n\varepsilon(1-\varepsilon)\vec{\nabla}p \qquad (16)$$

Identifying $KV = \gamma$, this last equation becomes,

$$\frac{\vec{j}_{eff}}{\rho_f} = -D_0 N_p \phi \vec{\nabla}\varepsilon + \varepsilon(1-\varepsilon)\vec{v}_D \qquad (17)$$

This is a generalized Darcy's law incorporating the effect of filling diffusion. For uniform filling (no diffusion) we recover, up to a constant factor depending on the filling factor, its usual form. This equation leads also to equilibrium configurations (vanishing current) characterized by the condition,

$$\frac{k}{\mu}p + \frac{D_0 N \phi}{2} \ln\left(\frac{\varepsilon}{1-\varepsilon}\right) = C \qquad (18)$$

Where $C$ denotes a constant. This condition is equivalent to a condition of constant chemical potential [13] throughout the medium. The equilibrium filling factor (at any point) finally reads,

$$\varepsilon = \frac{1}{1+\exp\left(\frac{2k(p-p^*)}{\mu D_0 N\phi}\right)} \tag{19}$$

It can be regarded as a (local) state equation. The filling parameter adopts a remarkable form: it behaves as a Fermi-Dirac function [13]. The threshold pressure $p^*$ is imposed by the constant value in Eq. 18 and the network features. The dual filling $1-\varepsilon$ (emptiness) is mathematically equivalent to a filling associated with the pressure reversal $(p(p^*) \to -p(p^*))$: the pores fullness and emptiness are thus related by a discrete transformation similar to that connecting particles and holes in quantum theory [13]. This connection will be referred to as duality symmetry. Indeed, the transformations $\varepsilon \to 1-\varepsilon$ and $p \to -p$ ($\vec{v}_D \to -\vec{v}_D$) preserve the dynamical Eq. 15. As the pressure reversal is equivalent to the Darcy velocity field reversal, this symmetry shouldn't be confused with the usual time reversal invariance (which does not obey Eq. 15 as a diffusion equation). Further studies to explore the richness of that model are actually in progress, especially in order to build up a possible framework of thermodynamics of such complex media incorporating their topological features and the influence of external deformations on the flow properties.

## 5. Applications

In the present section, potential applications of our approach of porous media to different fundamental or practical situations are discussed.

Lubricating films [14] flowing between rough solids in contact is a practical situation encountered in many industrial applications. The fluid is inserted into the voids separating the contacting solids (boundary lubrication). In the relevant case of the so-called mixed regime, the interface clearly resembles a disordered porous structure with a low thickness (quasi-2D film). The mixed regime is of special importance in industrial applications since it corresponds, according to the Stribeck curve [15], to a low friction coefficient. In that field, many issues remain unanswered or fuzzy. Our approach can help assessing the friction coefficient through its connection with the filling parameter variations over the interface, as well as the fluid volume trapped between the solids. This last parameter is related to the fluid film thickness, which partly determines the flowing regime.

On the fundamental side, another problem of peculiar interest for our group, regards intracranial dynamics and its pathologies [17]. This difficult problem has been tackled in many ways, as can be noticed in the available literature [18,19]. But these studies always have to face the unbelievable complexity of the intracranial system, a complexity with both structural and functional sides. An important question regards the description of the brain deformation dynamics, coupled to the intracranial fluids (cerebrospinal fluid (CSF) and blood dynamics), and its influence on the overall intracranial dynamics. More especially, that knowledge would shed light on the mechanism of related conditions such as hydrocephalus [17,20,21]. This condition is often depicted as a disorder of CSF hydrodynamics. The peripheral zones of brain are known to partly absorb CSF: brain matter then acts as a porous visco-elastic structure [17]. CSF diffusion through brain matter can be approached with our model to assess CSF pressure within. But as opposed to solids, brain is an easily deformable medium, and this affects deeply the CSF flowing regime (strong coupling). More especially, due to arterial blood pulses, the CSF pressure is modulated, resulting in dynamical flows of the Womersley type. We can thus expect that this application should generate new interesting features in the field and improve the ability of our model to account for the elastic properties of the solid matrix.

## 6. Conclusion

We have proposed a new theoretical way to tackle the complexity of porous media incorporating many of their features such as disorder effects, topology of the pores network, flowing regimes of the trapped fluid. Based on a discrete stochastic description of the network, it leads to an effective continuum description derived from the "microscopic" equation through a straightforward procedure. The most fundamental consequences of the model regard the emergence of Darcy's law and its connection to the network topology and disorder of the pores system and on the other side, the prediction of full/empty pores duality symmetry arising from the coupling between the pores filling dynamics and the pressure dynamics. A typical signature of that coupling is the prediction of steady out of equilibrium filling states and its dependence upon pressure which appears to follow a Fermi-Dirac type law. Application of this stochastic approach to fundamental or industrial problems involving porous media comprising a fluid component is considered, having in view to shed new light on these problems and improve the model.